\documentclass[notitlepage,nofootinbib,floatfix,amsmath,amssymb,showpacs,superscriptaddress]{revtex4-1}

\usepackage{graphicx}
\usepackage{dcolumn}
\usepackage{bm}
\usepackage{color}
\def\xslash{x\!\!\!\slash }
\def\qslash{q\!\!\!\slash }

\begin{document}

\title{Tensor form factors of nucleon in QCD}

\author{G\"{u}ray Erkol}
\affiliation{Laboratory for Fundamental Research, Ozyegin University, Kusbakisi Caddesi No:2 Altunizade, Uskudar Istanbul 34662, Turkey}
\email{guray.erkol@ozyegin.edu.tr}
\author{Altug Ozpineci}%
\affiliation{Physics Department, Middle East Technical University, 06531 Ankara, Turkey}
\email{ozpineci@metu.edu.tr}
\date{\today}

\begin{abstract}
We extract the isovector tensor nucleon form factors, which play an important role in understanding the transverse spin structure of the nucleon when related to the quark helicity-flip generalized parton distributions via their first moments. We employ the light-cone QCD sum rules to leading order in QCD and include distribution amplitudes up to twist 6 in order to calculate the three tensor form factors $H_T$, $E_T$ and $\tilde{H}_T$. Our results agree well with those from other approaches in the low and high momentum-transfer regions.
\end{abstract}
\maketitle

At the twist-two level the quark structure of the nucleon can be described by three quark distribution functions $f^q_1(x)$, $g^q_1(x)$ and $h^q_1(x)$, where $q$ is the corresponding quark. The \emph{unpolarized distribution}, $f^q_1(x)$, and the spin-dependent \emph{helicity distribution}, $g^q_1(x)$, give a measure of quark contribution to longitudinal momentum and the net helicity of a nucleon. These two chiral-even distribution functions are well-known quantities, which can be extracted from inclusive deep-inelastic scattering data. The \emph{transversity distribution}, $h^q_1(x)$, measures the net density of quarks having parallel polarization to that of the transversely polarized nucleon. $h^q_1(x)$, being a chiral-odd quantity, is difficult to measure experimentally because it requires a chiral-odd probe to couple in the cross section. The transversely polarized Drell-Yan process~\cite{Jaffe:1991kp, *Jaffe:1991ra, Cortes:1991ja} and electro- and photo-production of mesons off the polarized nucleons~\cite{Ivanov:2002jj, *Enberg:2006he, *Beiyad:2010cxa} are suggested as appropriate ways to measure $h^q_1(x)$. Recently, $h^q_1(x)$ has been extracted for the first time by an analysis of the experimental data from Belle~\cite{Abe:2005zx}, HERMES~\cite{Airapetian:2004tw} and COMPASS~\cite{Ageev:2006da} collaborations on azimuthal single spin asymmetry in semi-inclusive deep-inelastic scattering~\cite{Anselmino:2007fs, *Anselmino:2008jk, *Bacchetta:2011ip}. 

At the leading twist, a more general description of the nucleon structure is given by eight generalized parton distributions (GPD): Two chiral-even spin-independent distributions, $H$ and $E$, two chiral-even spin-dependent distributions, $\tilde{H^q}$ and $\tilde{E^q}$ and four chiral-odd spin-dependent distributions, $H^q_T$, $E^q_T$, $\tilde{H}^q_T$ and $\tilde{E}^q_T$~\cite{Ji:1996ek, *Radyushkin:1997ki, *Hoodbhoy:1998vm, *Diehl:2001pm}. Important information about the quark-gluon structure of the nucleon is encoded in these quantities. They describe, \emph{e.g.}, how partons are distributed in the transverse plane with respect to motion of the nucleon or the contribution of quark orbital angular momentum to total spin of the nucleon (see, \emph{e.g.}, the reviews in Refs.~\cite{Diehl:2003ny, *Ji:2004gf}).

Taking the moments of the GPDs one can obtain the nucleon form factors of the local currents. The first moments of the helicity-conserving GPDs, $H^q$, $E^q$ and $\tilde{H}^q$, $\tilde{E}^q$, are constrained by the nucleon form factors of the electromagnetic and axial currents, respectively. Similarly, the helicity-flip GPDs, $H^q_T$, $E^q_T$, $\tilde{H}^q_T$ and $\tilde{E}^q_T$, can be related to the nucleon form factors of the tensor current by taking the first moment. More explicitly, we have
\[
\int_{-1}^{1} dx~X(x,\xi,q^2)=X(q^2),
\]
for $X=H^q_T$, $E^q_T$ and $\tilde{H}^q_T$. Note that the first moment of $\tilde{E}^q_T(x,\xi,q^2)$ vanishes therefore we have $\tilde{E}^q_T(q^2)=0$~\cite{Hagler:2004yt}. Here $q=p^\prime-p$ is the momentum transfer and $\xi=-n\cdot q/2$ denotes the longitudinal momentum transfer, where $n$ is a light-like vector. The nucleon matrix elements of the tensor isovector current are parameterized in terms of above form factors as follows:
\begin{equation}
	\langle N(p^\prime)|T_{\mu\nu}|N(p)\rangle=\bar{u}(p^\prime)\left[i\sigma_{\mu\nu} H^{I=1}_T(q^2) +\frac{\gamma_\mu q_\nu-\gamma_\nu q_\mu}{2 m_N} E^{I=1}_T(q^2)+\frac{P_\mu q_\nu-P_\nu q_\mu}{2 m_N^2} \tilde{H}^{I=1}_T(q^2) \right] u(p),
\end{equation}
where $T_{\mu\nu}=\bar{u}i\sigma_{\mu\nu} u-\bar{d}i\sigma_{\mu\nu} d$ is the isovector tensor current, $P=p^\prime+p$, $m_N$ is the nucleon mass, and $X^{I=1}\equiv X^u - X^d$ for any of the form factors, $X=H_T,~E_T$ or $\tilde H_T$ (hereafter, the $I=1$ superscript will be dropped since we will be dealing with the $I=1$ case exclusively).  The tensor form factors at $Q^2=-q^2=0$ give what are called the \emph{tensor charges}. Previously, they have been investigated in the framework of QCD sum rules~\cite{He:1994gz, He:1996wy}, relativistic quark models~\cite{Schmidt:1997vm, *Pasquini:2005dk}, axial-vector dominance model~\cite{Gamberg:2001qc} and chiral-quark soliton model($\chi$QSM)~\cite{Kim:1995bq, *Kim:1996vk, *Lorce:2007fa}. The dependence of the tensor form factors on $Q^2$ have been extracted up to 1~GeV$^2$ from $\chi$QSM~\cite{Ledwig:2010tu} and up to a few GeV$^2$ from lattice QCD~\cite{Gockeler:2005cj,Hagler:2007xi}.

Our aim in this work is to study the nucleon tensor form factors in the framework of light-cone QCD sum rules method (LCSR)~\cite{Shifman:1978bx, *Shifman:1978by, *Reinders:1984sr, *Ioffe:1983ju, *Braun:1988qv, *Balitsky:1989ry, *Chernyak:1990ag}. LCSR have proved to be rather successful in extracting the values of the hadron form factors at high-momentum transfers, \emph{e.g.}, the electromagnetic and the axial form factors of the nucleon have been calculated to leading order and with higher-twist corrections in Refs.~\cite{Braun:2006hz, Erkol:2011qh, *Aliev:2007pi}. To our knowledge, while various estimations of $H_T(Q^2)$ at low- and high-momentum transfers exist, with this study it is the first time in the literature $E_T(Q^2)$ and $\tilde{H}_T(Q^2)$ are extracted.

In this method one starts with the following two-point correlation function:
\begin{equation}\label{corrf}
	\Pi^{N}_{\mu\nu}(p,q)=i\int d^4 x e^{iqx} \langle 0 |T[\eta_N(0)T_{\mu\nu}(x)]|N(p)\rangle,
\end{equation}
where $\eta_N(x)$ is the nucleon interpolating field. There are several local operators with the quantum numbers of spin-1/2 baryons one can choose from. Here we work with the general form of the nucleon interpolating field parameterized as follows:
\begin{equation}\label{intf}
	\eta_N=2\epsilon^{abc}\sum_{\ell=1}^{2}(u^{aT}(x) C J_1^\ell d^b(x))J_2^\ell u^c(x),
\end{equation} 
with $J_1^1=I$, $J_1^2=J_2^1=\gamma_5$ and $J_2^2=\beta$, which is an arbitrary parameter that fixes the mixing of two local operators. We would like to note that when the choice $\beta=-1$ is made the interpolating field above gives what is known as Ioffe current for nucleon. Here $u(x)$ and $d(x)$ denote the $u$- and $d$-quark fields, respectively, $a$, $b$, $c$ are the color indices and $C$ denotes charge conjugation.

The short-distance physics corresponding to high momenta $p^{\prime 2}$ and $q^2$ is calculated in terms of quark and gluon degrees of freedom. Inserting the interpolating field in Eq.~\eqref{intf} into the correlation function in Eq.~\eqref{corrf}, we obtain
\begin{widetext}
\begin{align}\label{corrfunc}
	\begin{split}
	\Pi^N_{\mu\nu}=&\frac{1}{2}\int d^4 x e^{iqx}\sum^2_{\ell=1}\left\{(C J_1^\ell)_{\alpha\gamma} \left[J_2^\ell S_u(-x)i \sigma_{\mu\nu}\right]_{\rho\beta}4\epsilon^{abc}\langle 0|u_{\alpha}^a(0) u_{\beta}^b(x) d_{\gamma}^c(0)|N\rangle\right.\\
	&+  (J_2^\ell)_{\rho\alpha}\left[(CJ_1^\ell)^T S_u(-x)i \sigma_{\mu\nu}\right]_{\gamma\beta}4\epsilon^{abc}\langle 0|u_{\alpha}^a(x) u_{\beta}^b(0) d_{\gamma}^c(0)|N\rangle\\
	&\left. +  (J_2^\ell)_{\rho\beta}\left[CJ_1^\ell S_d(-x)i \sigma_{\mu\nu}\right]_{\alpha\gamma}4\epsilon^{abc}\langle 0|u_{\alpha}^a(0) u_{\beta}^b(0) d_{\gamma}^c(x)|N\rangle\right\},
	\end{split}
\end{align}
\end{widetext}
where $u$ and $d$ denote the quark fields. $S_{u,d}$ represent the quark propagator
\begin{equation}\label{qprop}
	S_q(x)=\frac{i\xslash}{2\pi^2x^4}-\frac{\langle q\bar{q}\rangle}{12}\left(1+\frac{m_0^2 x^2}{16}\right)-ig_s\int^1_0 d\upsilon\left[\frac{\xslash}{16\pi^2x^4} G_{\mu\nu}\sigma^{\mu\nu}-\upsilon x^\mu G_{\mu\nu}\gamma^\nu\ \frac{i}{4\pi^2x^2}\right].
\end{equation}
Here the first term gives the hard-quark propagator. The second term represents the contributions from the nonperturbative structure of the QCD vacuum, namely, the quark and quark-gluon condensates. These contributions are removed by Borel transformations. The last term is due to the correction in the background gluon field and gives rise to four-particle ($N=4$ with an additional gluon) and five-particle ($N=5$ with an additional quark--anti-quark pair) baryon distribution amplitudes. In Ref.~\cite{Diehl:1998kh}, higher Fock states ($N>3$) of the nucleon wavefunction have been considered and it has been shown by an application to electromagnetic form factors of the nucleon that the valence Fock states dominate while $N=4,5$ Fock states give only small contributions. Following the common practice, in this work we shall not take into account such contributions, which leaves us with the first  term in Eq.~\eqref{qprop} to consider.

The matrix elements of the local three-quark operator 
\begin{eqnarray} 
4\epsilon^{abc}\langle 0|u_{\alpha}^a(a_1 x) u_{\beta}^b(a_2 x) d_{\gamma}^c(a_3 x)|N\rangle
\label{mel}
\end{eqnarray}
($a_{1,2,3}$ are real numbers in the interval $[0,1]$ denoting the position of the  quarks along the ray pointing from the origin to the space-time point $x$) can be expanded in terms of DAs using the Lorentz covariance, the spin and the parity of the baryon. Based on a conformal expansion using the approximate conformal invariance of the QCD Lagrangian up to 1-loop order, the DAs are then decomposed into local nonperturbative parameters, which can be estimated using QCD sum rules or fitted so as to reproduce experimental data. 

The long-distance side of the correlation function is obtained using the analyticity of the correlation function, which allows us to write the correlation function in terms of a dispersion relation of the form
\[\Pi^N_{\mu}(p,q)=\frac{1}{\pi}\int_0^\infty \frac{\text{Im}\Pi^N_\mu(s)}{(s-p^{\prime 2})}ds\]
The ground-state hadron contribution is singled out by utilizing the zero-width approximation
\[\text{Im}~\Pi^N_\mu=\pi \delta(s-m_N^2)\langle 0|\eta^N|N(p^\prime)\rangle \langle N(p^\prime)|T_{\mu\nu}|N(p)\rangle\]
expressing the correlation function as a sharp resonance plus continuum with the continuum threshold, $s_0$. The matrix element of the interpolating current between the vacuum and baryon state is defined as
\[\langle 0|\eta^N|N(p,s)\rangle=\lambda_N\upsilon(p,s)\]
where $\lambda_N$ is the nucleon overlap amplitude and $\upsilon(p,s)$ is the nucleon spinor.

The QCD sum rules are obtained by matching the short-distance calculation of the correlation function with the long-distance calculation. Using the most general decomposition of the matrix element in Eq. (\ref{mel})  we obtain the following sum rules for the tensor form factors:
	\begin{align}
		\begin{split}
			H_T=&\frac{1}{4\lambda_N}e^{m_N^2/M^2}\left\{m_N Y(1,1)+\sum_{i=1}^{2} \frac{m_N^3}{M^2} Y(2,4i+1) +\frac{m_N^3}{M^2} Y(1,13) +\sum_{i=1}^{2} m_N^3 Z(4i+1) + m_N^3 x_0 Z(13)  \right\},\\
			E_T=&\frac{1}{2\lambda_N}e^{m_N^2/M^2}\left\{m_N Y(1,1)+\sum_{i=1}^{2} \frac{m_N^3}{M^2} Y(2,4i+1)+\frac{m_N^3}{M^2} Y(1,13) +\sum_{i=1}^{2} m_N^3 Z(4i+1)+m_N^3 x_0Z(13) \right\},\\
			\tilde{H}_T=&\frac{1}{2\lambda_N}e^{m_N^2/M^2}\left\{m_N Y(1,1)+\sum_{i=1}^{2} \frac{m_N^3}{M^2} Y(2,4i+1) +\frac{m_N^3}{M^2} Y(1,13) +\sum_{i=1}^{2} m_N^3 Z(4i+1)+m_N^3 x_0 Z(13) \right\},\\
		\end{split}
	\end{align}
where we define
	\begin{align}
		\begin{split}
	Y(k,n)&=\int^1_{x_0}\frac{dt_2}{(t_2)^k}e^{-s(t_2)/M^2}\left\{\left[(1-\beta)F_n+(1+\beta)F_{n+1}\right]+ \left[(1-\beta)F_{n+2}+(1+\beta)F_{n+3}\right]\right\},\\
		Z(n)&=\frac{e^{-s_0/M^2}}{Q^2+x_0^2 m_N^2} \left[(1-\beta)(F_n+F_{n+2}) +(1+\beta)(F_{n+1}+F_{n+3})\right],
	\end{split}
\end{align}
for $H_T$ form factor at the structure $i\qslash\sigma_{\mu\nu}$, for $E_T$ form factor at the structure $\gamma_\mu q_\nu-\gamma_\nu q_\mu$ and for $\tilde{H}_T$ form factor at the structure $P_\mu q_\nu-P_\nu q_\mu$. The explicit form of the functions that appear in the above sum rules are given in terms of DAs as follows:  \\
for $H_T$: 
{\allowdisplaybreaks
	\begin{align*}
		&F_1=2\int_0^{1-t_2} dt_1 \left[-P_1+T_1-T_2+T_7\right](t_1,t_2,1-t_1-t_2),\\
		&F_2=\int_0^{1-t_2} dt_1 \left[-A_1+A_2+2A_3+V_1-V_2+2V_3\right](t_1,t_2,1-t_1-t_2),\\
		&F_3=2\int_0^{1-t_2} dt_1 \left[-A_1+A_3-V_1-V_3\right](t_1,1-t_1-t_2,t_2),\\
		&F_4=2\int_0^{1-t_2} dt_1 \left[-P_1+S_1-2T_1-T_3-T_7\right](t_1,1-t_1-t_2,t_2),\\
		&F_5=0,\\
		&F_6=0,\\
		&F_7=\int_0^{1-t_2} dt_1 \left[A_1^M+V_1^M\right](t_1,1-t_1-t_2,t_2),\\
		&F_8=\int_0^{1-t_2} dt_1 \left[3T_1^M\right](t_1,1-t_1-t_2,t_2),\\
		&F_9=-4\int_1^{t_2}d\lambda \int_1^\lambda d\rho \int_0^{1-\rho} dt_1 \left[T_1-T_2-T_5+T_6-2T_7-2T_8\right](t_1,\rho,1-t_1-\rho),\\
		&F_{10}=4\int_1^{t_2}d\lambda \int_1^\lambda d\rho \int_0^{1-\rho} dt_1 \left[A_1-A_2+A_3+A_4-A_5+A_6-V_1+V_2+V_3+V_4+V_5-V_6\right](t_1,\rho,1-t_1-\rho),\\
		&F_{11}=\int_1^{t_2}d\lambda \int_1^\lambda d\rho \int_0^{1-\rho} dt_1 \left[A_1-A_2+A_3+A_4-A_5+A_6+V_1-V_2-V_3-V_4-V_5+V_6\right] (t_1,1-t_1-\rho,\rho),\\
		&F_{12}=-\int_1^{t_2}d\lambda \int_1^\lambda d\rho \int_0^{1-\rho} dt_1 \left[-T_1+3T_2-2T_3-2T_4+3T_5-T_6+4T_7+4T_8\right] (t_1,1-t_1-\rho,\rho),\\
		&F_{13}=0,\\
		&F_{14}=0,\\
		&F_{15}=0,\\
		&F_{16}=0,\\
	\end{align*}
}
for $E_T$: 
{\allowdisplaybreaks
	\begin{align*}
		&F_1=2\int_0^{1-t_2} dt_1 \left[-P_1+S_1+2T_1-T_3-T_7\right](t_1,t_2,1-t_1-t_2),\\
		&F_2=2\int_0^{1-t_2} dt_1 \left[-A_1+V_1\right](t_1,t_2,1-t_1-t_2),\\
		&F_3=2\int_0^{1-t_2} dt_1 \left[-A_1+A_2-V_1+V_2\right](t_1,1-t_1-t_2,t_2),\\
		&F_4=-4\int_0^{1-t_2} dt_1 \left[T_1\right](t_1,1-t_1-t_2,t_2),\\
		&F_5=-2\int_0^{1-t_2} dt_1 \left[T_1^M\right](t_1,t_2,1-t_1-t_2),\\
		&F_6=-2\int_0^{1-t_2} dt_1 \left[-A_1^M+V_1^M\right](t_1,t_2,1-t_1-t_2),\\
		&F_7=0,\\
		&F_8=4\int_0^{1-t_2} dt_1 \left[T_1^M\right](t_1,1-t_1-t_2,t_2),\\
		&F_9=2\int_1^{t_2}d\lambda \int_1^\lambda d\rho \int_0^{1-\rho} dt_1 \left[-T_1+T_2+T_5-T_6+2T_7+2T_8\right](t_1,\rho,1-t_1-\rho),\\
		&F_{10}=0,\\
		&F_{11}=2\int_1^{t_2}d\lambda \int_1^\lambda d\rho \int_0^{1-\rho} dt_1 \left[A_1-A_2+A_3+A_4-A_5+A_6+V_1-V_2-V_3-V_4-V_5+V_6\right] (t_1,1-t_1-\rho,\rho),\\
		&F_{12}=-4\int_1^{t_2}d\lambda \int_1^\lambda d\rho \int_0^{1-\rho} dt_1 \left[T_2-T_3-T_4+T_5+T_7+T_8\right] (t_1,1-t_1-\rho,\rho),\\
		&F_{13}=2\int_{t_2}^{1} dt_1\int_0^{1-t_1} d\rho  \left[P_1-P_2+S_1-S_2-2T_1-2T_2+T_3+T_4+2T_5+3T_7-T_8 \right](t_1,1-t_1-\rho,\rho),\\
		&F_{14}=-2 \int_{t_2}^{1} dt_1\int_0^{1-t_1} d\rho\left[A_1+A_2+2A_3-A_4-2A_5+V_1+V_2-2V_3+V_4-2V_5 \right](t_1,1-t_1-\rho,\rho),\\
		&F_{15}=-2 \int_{t_2}^1 d\rho  \int_0^{1-\rho} dt_1 \left[A_1-A_2+A_3-V_1+V_2+V_3 \right](t_1,1-t_1-\rho,\rho),\\
		&F_{16}=2\int_{t_2}^1 d\rho  \int_0^{1-\rho} dt_1\left[P_1-P_2-S_1+S_2+2T_1-T_3-T_4-T_7-T_8 \right](t_1,1-t_1-\rho,\rho),\\
	\end{align*}
}
and for $\tilde{H}_T$: 
{\allowdisplaybreaks
	\begin{align*}
		&F_1=0,\\
		&F_2=\int_0^{1-t_2} dt_1 \left[V_1-A_1\right](t_1,t_2,1-t_1-t_2),\\
		&F_3=0,\\
		&F_4=-2\int_0^{1-t_2} dt_1 \left[T_1\right](t_1,1-t_1-t_2,t_2),\\
		&F_5=0,\\
		&F_6=-\int_0^{1-t_2} dt_1 \left[V_1^M-A_1^M\right](t_1,t_2,1-t_1-t_2),\\
		&F_7=0,\\
		&F_8=2\int_0^{1-t_2} dt_1 \left[T_1^M\right](t_1,1-t_1-t_2,t_2),\\
		&F_9=0,\\
		&F_{10}=0,\\
		&F_{11}=0,\\
		&F_{12}=-2\int_1^{t_2}d\lambda \int_1^\lambda d\rho \int_0^{1-\rho} dt_1 \left[T_2-T_3-T_4+T_5+T_7+T_8\right] (t_1,1-t_1-\rho,\rho),\\
		&F_{13}=\int_{t_2}^{1} dt_1\int_0^{1-t_1} d\rho  \left[P_1-P_2+S_1-S_2-4T_1-2T_2+3T_3+T_4+2T_5+5T_7-T_8 \right](t_1,1-t_1-\rho,\rho),\\
		&F_{14}= \int_{t_2}^{1} dt_1\int_0^{1-t_1} d\rho  \left[A_1+A_2+2A_3-A_4-2A_5+V_1+V_2-2V_3+V_4-2V_5 \right](t_1,\rho,1-t_1-\rho),\\
		&F_{15}=2 \int_{t_2}^1 d\rho  \int_0^{1-\rho} dt_1 \left[A_1-A_2+A_3-V_1+V_2+V_3 \right](t_1,1-t_1-\rho,\rho),\\
		&F_{16}= - \int_{t_2}^1 d\rho  \int_0^{1-\rho} dt_1 \left[P_1-P_2-S_1+S_2+2T_1-T_3-T_4-T_7-T_8 \right](t_1,1-t_1-\rho,\rho).\\
	\end{align*}
}
The DA's $P_i$, $A_i$, $S_i$ and $T_i$ appear in the most general decomposition of the nucleon matrix element~\cite{Braun:2000kw} and the twist-5 DA's $A_1^M$, $V_1^M$and $T_1^M$ have been calculated in Ref.~\cite{Braun:2006hz}. In obtaining the sum rules, Borel transformation is applied to eliminate the subtraction terms in the spectral representation of the correlation function and to exponentially suppress the contributions from excited and continuum states. The contributions of the higher states and the continuum are modeled using the quark-hadron duality and subtracted. Both of the Borel transformation and the subtraction of the higher states are carried out using the following substitution rules:
\begin{align}
	\begin{split}
		&\int dx \frac{\rho(x)}{(q-xp)^2}\rightarrow -\int_{x_0}^1\frac{dx}{x}\rho(x) e^{-s(x)/M^2},\\
		&\int dx \frac{\rho(x)}{(q-xp)^4}\rightarrow \frac{1}{M^2} \int_{x_0}^1\frac{dx}{x^2}\rho(x) e^{-s(x)/M^2}+\frac{\rho(x)}{Q^2+x_0^2 m_N^2} e^{-s_0/M^2},
	\end{split}
\end{align}
where
\[s(x)=(1-x)m_N^2+\frac{1-x}{x}Q^2.\]
$M$ is the Borel mass and $x_0$ is the solution of the quadratic equation for $s=s_0$:
\[x_0=\left[\sqrt{(Q^2+s_0-m_N^2)^2+4m_N^2(Q^2)}-(Q^2+s_0-m_N^2)\right]/(2m_N^2),\] where $s_0$ is the continuum threshold.

To obtain a numerical prediction for the form factors, the residue,
$\lambda_N$, is also required. The residue can be obtained from the mass sum rule, which is given by:
\begin{equation}
\label{residue}
\lambda_{N}^2 e^{-m_{N}^2/M^2} =
{M^6\over 1024 \pi^2} (5 + 2 \beta + 5 \beta^2) E_2(x) - {m_0^2\over 96 M^2} (-1+\beta)^2 \langle 
\bar{q} q \rangle^2 - {m_0^2\over 8 M^2} (-1+\beta^2) \langle  \bar{q} q \rangle^2 
+ {7\over 24}  (-1+\beta^2) \langle  \bar{q} q \rangle^2 \nonumber , 
\end{equation}
where $x = s_0/M^2$, and
\begin{eqnarray}
\label{nolabel}
E_n(x)=1-e^{-x}\sum_{i=0}^{n}\frac{x^i}{i!}~. \nonumber
\end{eqnarray}
We use the following parameter values: $\langle \bar{q}q\rangle=-(0.243)^3$~GeV$^3$, $m_0^2=0.8$~GeV$^2$ and $m_N=0.94$~GeV.

The DAs of the nucleon are given in Ref.~\cite{Braun:2006hz} and they are expressed in terms of some nonperturbative parameters which are calculated using QCD sum rules or phenomenological models. In this work, we use Chernyak-Zhitnitsky-like model of the DAs (see Ref.~\cite{Braun:2006hz} for details) to calculate the tensor form factors. Note that unlike other form factors such as axial ones, the tensor form factors are renormalization-scale, $\mu$, dependent~\cite{He:1994gz}. Our results correspond to $\mu^2=1$~GeV$^2$ as we are using the numerical values of the DAs at this scale~\cite{Chernyak:1984bm}.

There are several parameters in the sum rules that need to be determined. We use the continuum threshold value $s_0\sim$ 2.25~GeV$^2$, which is pretty much fixed in the literature from nucleon spectrum analysis. We also give our numerical results at $M^2=$2~GeV$^2$. The sum rules and hadron properties are expected to be independent of the mixing parameter $\beta$. In order to see if we can achieve such an independence, we make a reparameterization using $\beta=\tan\theta$ and observe that a stability region with respect to a change in the mixing parameter can be found around $\cos\theta\sim$0, which we use to produce our results.

\begin{figure}[t]
	\includegraphics[width=\textwidth]{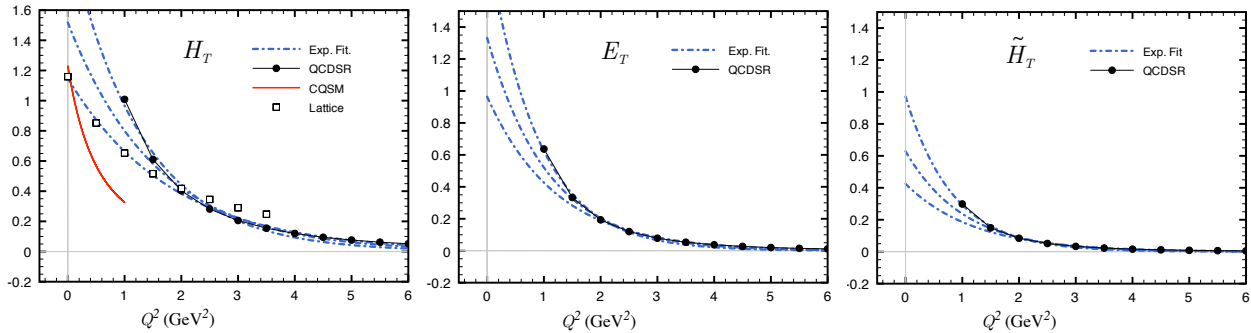}
	\caption{\label{tensor_all} Tensor form factors $H_T(Q^2)$, $E_T(Q^2)$ and $\tilde{H}_T(Q^2)$ as a function of $Q^2$ above $Q^2=1$~GeV$^2$. We also show our fitted form factors (see the text for details) and give the results for $H_T$ as obtained from $\chi$QSM~\cite{Ledwig:2010tu} and from lattice QCD~\cite{Gockeler:2005cj}}
\end{figure}

We plot the tensor form factors $H_T(Q^2)$, $E_T(Q^2)$ and $\tilde{H}_T(Q^2)$ as a function of $Q^2$ in Fig.~\ref{tensor_all}. We show the form factors above $Q^2=1$~GeV$^2$. We also give the results for $H_T$ as obtained from $\chi$QSM~\cite{Ledwig:2010tu} and from lattice QCD~\cite{Gockeler:2005cj}, which have been calculated at a renormalization scale of $\mu^2=0.36$~GeV$^2$ and $\mu^2=4$~GeV$^2$, respectively. We use the following relation~\cite{Gluck:1994uf, *Barone:2001sp} to relate these results of the form factors to those at $\mu^2=1$~GeV$^2$:
\begin{equation}
	X(\mu^2)=\left(\alpha_s(\mu^2)\over\alpha_s(\mu_i^2)\right)^{4/27}\left[1-\frac{337}{468\pi}[\alpha_s(\mu_i^2)-\alpha_s(\mu^2)]\right]X(\mu_i^2),
\end{equation}
where $\mu_i$ is the initial renormalization scale and 
\begin{equation}
	\alpha_s(\mu^2)=\frac{4\pi}{9\ln(\mu^2/\Lambda_\text{QCD}^2)}\left[1-\frac{64}{81}\frac{\ln\ln(\mu^2/\Lambda_\text{QCD}^2)}{\ln(\mu^2/\Lambda_\text{QCD}^2)}\right]
\end{equation}
with $\Lambda_\text{QCD}=0.250$~GeV and $N_c=N_f=3$.

In order to extrapolate the form factors to low-momentum region, we have fitted QCDSR results to various forms including monopole and dipole, which unfortunately fail to give a reasonable description of data with a two-parameter fit. On the other hand, an exponential form, {\it viz.},
\begin{equation}\label{expform}
	F_T(Q^2)=F_{T}(0) \exp[-Q^2/m_{T}^2],
\end{equation}
allows a plausible description of data with a two-parameter fit. Note that the LCSR are reliable for $Q^2$ larger than a few GeV$^2$. We have, however, tried three fit regions, namely, $Q^2>1$~GeV$^2$, $Q^2>1.5$~GeV$^2$ and  $Q^2>2$~GeV$^2$. Our results are given in Table~\ref{fit_table}. $H_T(0)$ values are to be compared with those from experiment as $H_T(0)=0.75^{+0.18}_{-0.38} $\cite{Anselmino:2008jk}, from $\chi$QSM as $H_T(0)=1.23$~\cite{Ledwig:2010tu}, from external-field QCD sum rules method as $H_T(0)=1.29\pm0.51$~\cite{He:1996wy}, and from lattice QCD as $H_T(0)=1.15\pm0.15$~\cite{Gockeler:2005cj} at $\mu^2=1$~GeV$^2$. We find that our results for $H_T$ agree well with those from other theoretical approaches, when we make a fit from a more reliable high-$Q^2$ region, namely from $Q^2>2$~GeV$^2$ (all theoretical methods overestimate $H_T$ when compared to the phenomenological result in~\cite{Anselmino:2007fs, *Anselmino:2008jk}). In Fig.~\ref{tensor_all} we also show our fitted form factor as well as those from other theoretical approaches. In the high-$Q^2$ region we observe that our $H_T$ results compare well with lattice-QCD results. With the finding that an exponential fit above 2~GeV$^2$ gives more accurate results for $H_T$, we quote our final results for the other two form factors at $Q^2=0$ as extrapolated from this region as $E_T(0)=0.96$ and $\tilde{H}_T(0)=0.43$. 

In conclusion, we have extracted the isovector tensor form factors of nucleon, which play an important role in understanding the transverse spin structure of the nucleon when related to the quark helicity-flip generalized parton distributions via their first moments. To this end, we have employed LCSR method to leading order in QCD and included distribution amplitudes up to twist 6. We have found that our results for the $H_T$ form factor compare well with lattice-QCD results in the high-$Q^2$ region and with those from other theoretical approaches when extrapolated to low-$Q^2$ regions via an exponential fit. We have also calculated the tensor form factors $E_T$ and $\tilde{H}_T$ for the first time in the literature. We will extend our work with a detailed analysis of tensor form factors with respect to each quark contribution so as to extract isoscalar, octet and singlet form factors as well as those of strange baryons, in a future publication. 

\begin{table}[t]
	\caption{The values of exponential fit parameters, namely $F_T(0)$ and $m_{T}$, of tensor form factors. We give the results of the fits from three regions. }
	\addtolength{\tabcolsep}{2pt}
\begin{tabular}{ccccccc}
		\hline\hline 
		Form Factor & Fit Region~(GeV$^2$) & $F_{T}(0)$ & $m_{T}$~(GeV)& \\[0.5ex]
		\hline 
		& [2.0-10] & 1.15 & 1.35  & \\
		$H_T$ & [1.5-10]& 1.52 & 1.25  &  \\
		& [1.0-10] & 2.11 & 1.13  &  \\[1ex]
		& [2.0-10] & 0.96 & 1.11  &  \\
		$E_T$ & [1.5-10]& 1.33 & 1.03  &  \\
		& [1.0-10] & 1.92 & 0.94  &  \\[1ex]
		& [2.0-10] & 0.43 & 1.10  &  \\
		$\tilde{H}_T$ & [1.5-10]& 0.63 & 1.01  &  \\
		& [1.0-10] & 0.97 & 0.91  &  \\
		\hline\hline
	\end{tabular}
	\label{fit_table}
\end{table}

\acknowledgments
This work has been supported by The Scientific and Technological Research Council of Turkey (T\"{U}B{\.I}TAK) under project number 110T245. The work of A. O. is also partially supported by the European Union (HadronPhysics2 project Study of strongly interacting matter).
%

\end{document}